\documentclass[useAMS,usenatbib,usedcolumn,usegraphicx]{mn2e}
\def\etal{{\it et\thinspace al.}\ }
\def\Fe17{Fe\,{\sc xvii}\ }
\usepackage[dvips]{graphics}
\usepackage[dvipsnames,usenames]{color}

\newcommand{\be}{\begin{equation}}
\newcommand{\ee}{\end{equation}}
\title{Electron distributions in X-Ray plasmas:
spectral diagnostics with the 3s/3d line ratio in Fe\ {XVII}}
\author[Guo-Xin Chen and Anil K. Pradhan]
       {Guo-Xin Chen$^1$ and Anil K. Pradhan$^2$\\
       $^1$ ITAMP, Harvard-Smithsonian Centre for Astrophysics, 60
       Garden St., Cambridge, MA 02138, USA,\\ $^2$ Department of Astronomy,
 The Ohio State University, Columbus, OH 43210, USA}
\date{Accepted  xxxxxx 
      Received xxxxxx;
      in original form xxxxxx}

\pagerange{\pageref{firstpage}--\pageref{lastpage}}
\pubyear{2004}

\def\LaTeX{L\kern-.36em\raise.3ex\hbox{a}\kern-.15em
    T\kern-.1667em\lower.7ex\hbox{E}\kern-.125emX}

\begin{document}

\maketitle

\label{firstpage}

\begin{abstract}
Efforts to benchmark astrophysical observations with X-ray laboratory
measurements have been stymied by observed and measured differences of
up to a factor of two in the ratio '3s/3d' of \Fe17 lines at
$\lambda\lambda$ $\sim17\AA$ and $\sim15 \AA$ respectively. 
Using the electron distribution function (EDF) as a new physical
parameter, we compute the \Fe17 line ratios and account for these differences.
Based on large-scale relativistic close
coupling calculations using the Breit-Pauli R-matrix method,
revealing the precise effect of resonances in collisional excitation, we
employ collisional-radiative models using cross
sections convolved with both the Gaussian and the Maxwellian EDF.
Comparison with astrophysical observations and laboratory measurements
demonstrates that (a) the 3s/3d line ratio depends not only on the EDF 
but also on the electron temperature/energy of the source,
(b) plasma conditions in experimental measurements and
astrophysical observations may be quite different, and (c)
 departure from a Maxwellian should manifest itself in, and be
used as a diagnostics of, particle distributions in plasmas.

\end{abstract}


\section{INTRODUCTION}
 
 Astrophysical X-ray sources cover a wide range of physical conditions in
objects as diverse as stellar coronae, active
galactic nuclei, X-ray binaries, supernovae, and afterglows from
gamma-ray bursts. In principle, spectral
diagnostics should provide accurate information on the particular plasma
conditions, particularly from high-resolution observations possible with
X-ray satellites Chandra and XMM-Newton \cite{can00,huen01,sab99}.
However, the diagnostics are contingent upon accurate theoretical models
calibrated against precise experimental measurements. Several spectral
lines from neon-like \Fe17 are prominent constituents in the soft
X-ray region $\sim$15--17\ \AA,
and have long been observed as potential diagnostics (Fig. 1).
 Recent electron beam ion trap (EBIT) experiments on
\Fe17 line intensity ratios at
the Lawrence Livermore National Laboratory (LLNL) \cite{bei02}, and at the
National Institute for Standards and Technology (NIST) \cite{lam00},
may establish these diagnostics, provided the experimental results,
theoretical calculations, and astrophysical observations can be interpreted
consistently.
But the diagnostic capabilities of the \Fe17 X-ray line ratios have so 
far been limited because the line-formation mechanism is not well understood,
 and all spectral modeling codes carry large uncertainties.

\begin{figure}
\centering
\includegraphics[width=\columnwidth,keepaspectratio]{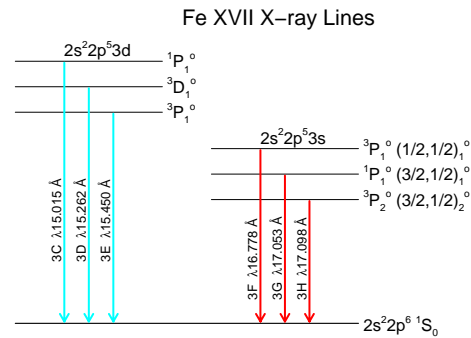}
\caption{Schematic \Fe17 transitions for three 3s X-ray lines (3F, 3G,
and 3H) and three 3d X-ray lines (3C, 3D, and 3E).
\label{plot1}
}
\end{figure}

A schematic representation of the lines of interest in this Letter 
is given in Fig.\,\ref{plot1}.
Excitation of the closed-shell $2p^6$ ground configuration of neon-like
\Fe17 to the $2p^5 \ 3s$ and $2p^5 \ 3d$ configurations corresponds to
sets of 3 lines each at $\sim17 \AA$ and $\sim15 \AA$, respectively
labeled as the '3s' and the '3d'. 
The interpretation of the 3s/3d ratios in \Fe17 from
current EBIT and tokamak experiments, and astrophysical observations,
are perplexing. Basically, there are two long-standing problems in \Fe17 
spectra: the so called 3C/3D ratio and the 3s/3d ratio in \Fe17
\cite{lam00,bei02}. The
3C/3D is the line intensity ratio of transitions labeled as
3C ($\lambda$\ 15.015\ {\AA}) and 3D ($\lambda$\ 15.262\ {\AA}).
The ratio 3s/3d=$\displaystyle{\frac{\rm I(3F+3G+3H)}{\rm I(3C+3D+3E)}}$
(line 3H is sometimes labeled as M2),
is the intensity ratio of three lines at $\lambda\lambda$\ 17.098, 17.053 and 16.778\ {\AA}
with one 3s electron in upper levels, over another set of three lines
at $\lambda\lambda$\ 15.015, 15.262 and 15.450\ {\AA} with one 3d electron in 
upper levels.

 In the past theoretical predictions of
the X-ray line intensity ratio 3C/3D were about 50\% or more higher than
astronomical observations or laboratory experiments.
Yet, the EBIT measurements were in agreement with each other and with 
observations.
This problem has been addressed in our recent
calculations \cite{che02,che03,che05} in the 
low electron impact energy region.
In this low energy region, we carried out ab initio calculations using
the Breit-Pauli {\sl R}-matrix (BPRM) method \cite{ber95} including
electron-impact excitation (EIE) resonances up to $n=4$ levels ($n$ is
the principal quantum number), which
play a key role in the solution of the 3C/3D problem. The
resonance enhancement is extrapolated from $n=4$ 
to $n=5$ for the higher electron-energy (temperature) region
(this extrapolation needs to be verified by further
calculations). Similarly, the 3s/3d problem pertains to
nearly a factor of 2 difference in the
intensity ratio between the two sets of EBIT measurements \cite{bei02,lam00}.
Moreover, there is very large scatter in the 3s/3d ratios from
all previous theoretical calculations and observations.
However, this case is more complicated than the 3C/3D problem for two reasons.
First, at low electron densities cascade effects have important or dominant
contributions to the formation of some or all 3s X-ray lines; while the 3C
and 3D lines are formed predominantly by EIE. Second, the 3s/3d ratio is
very sensitive to plasma conditions under which these X-ray lines are
formed. In contrast,
the 3C/3D ratio is not very source sensitive i.\,e.\ it is determined mainly
by atomic micro-processes. It is due to these two reasons that the
measured and observed 3s/3d ratios for given temperature and density may
differ by up to a factor of two --- a difference that is
a major unresolved problem in X-ray astrophysics.

One of main goals in this paper is also to
introduce electron distribution functions
(EDF) (e.\,g.\ Maxwellian and Gaussian EDF) as a new physical parameter
for the study of X-ray formation mechanism.
As evident from the results in this Letter, this new approach is essential
to solve the long-standing 3s/3d problem in Fe~{\small XVII}.
The main point is
that plasmas in EBITs and astrophysical sources may have different 
beam/particle distributions.
In some laboratory plasmas e.\,g.\ EBIT experiments, the electron beam
width may be Gaussian or non-Gaussian (non-Maxwellian).
In astrophysical objects on the other hand the EDF
is normally a Maxwellian, but there are important cases (such as
magnetically driven plasmas)
where the source may display a bi-Maxwellian character,
or a non-Maxwellian component of electron fluxes with
high-energy tails. The precise convolution of cross sections with
velocity distributions must therefore be generally source-specific.

     \section{Theory and computations}

      The extensive calculations of the \Fe17 collision strengths using
      the BPRM method and a 89-level close coupling expansion up to the
      $n$ = 4 levels is described in earlier works 
      \cite{che02,che03}.
      Resonance structures in the collision strengths are delineated;
       the positions, strengths, and distribution of resonances
      have been illustrated in detail for many of the transitions of
      interest in this work. 
      Considerable differences are found with respect to
     previous calculations that neglect resonances, such as several
     distorted wave results available in literature.  
       The energy dependence of the collision strengths due
      to resonances introduces a temperature dependence in the
      averaged collision strengths with temperature dependent EDF's such
      as a Maxwellian. Moreover, the many Rydberg series of resonances,
      converging on to most of the 89 levels in the wavefunction
      expansion, are not uniformly distributed in energy. Therefore a
      Gaussian average yields
      significantly different values, depending on the assumed energy
      widths, than a Maxwellian average.

In the present work, in order to address the issue of EDF influence on line
ratios in general, we recalculate the averaged collision
strengths, or rate coefficients, and reconstruct the
collisional-radiative model. 
We carry out separate convolutions of our detailed close-coupling cross
sections with specific beam distributions e.\,g.\ a Gaussian
average (GA) and a Maxwellian average (MA),
{\it before} the construction of the collisional-radiative model.
     The Maxwellian averaged rate
      coefficients are computed for all transitions as

 \be q_{ij}(T) = \frac{8.63 \times 10^{-6}{\rm cm}^3{\rm/s}\cdot
  \exp(-E_{ij}/kT_{\rm e})} {g_i \sqrt{T_{\rm e}/{\rm K}}}
   \Upsilon_{ij}(T_{\rm e}),\ee
    where g$_i$ is the statistical weight of the initial level and the
     quantity $\Upsilon_{ij}$ is the Maxwellian averaged collision
     strength:
      \be \Upsilon_{ij}(T) = \int_{E_j}^{\infty} \Omega_{ij}(E)
       \exp(-E/kT_{\rm e})\ {\rm d}(E/kT_{\rm e}). \ee

 In addition, calculation for all rate coefficients are repeated with
 Gaussian EDF's obtained using a
 specified energy spread $\Delta E$ to simulate different beam widths in EBIT
 experiments. The collisional-radiative
       model for the \Fe17 lines includes 3,916 transitions associated
             with the atomic model, which also employs new allowed and forbidden
	           transition probabilities for \Fe17 recently
		   computed by Nahar \etal (2003).

\section{Results and discussion}

\begin{figure*}
\begin{minipage}{148mm} % figure using both columns
\centering
\includegraphics[width=148mm,keepaspectratio]{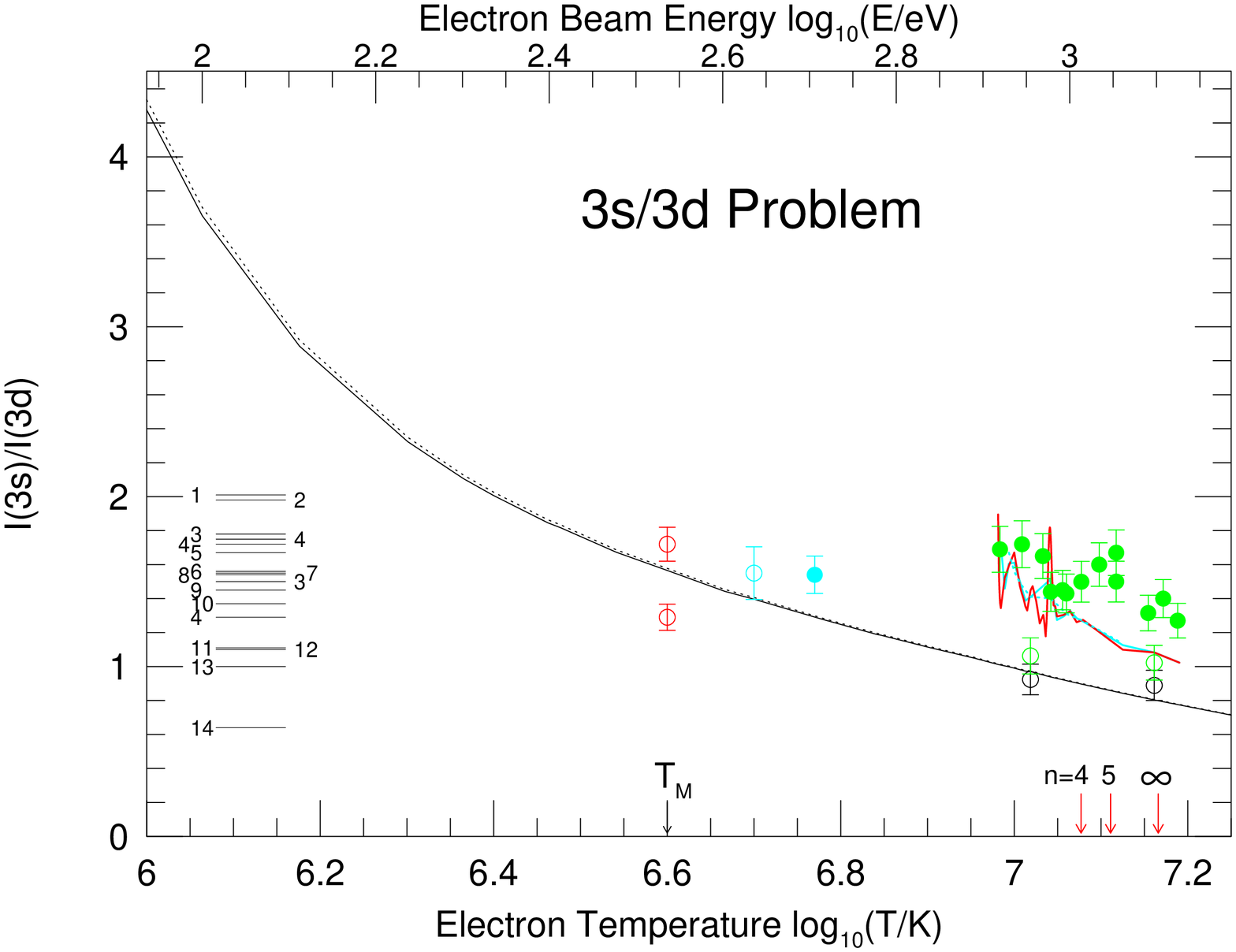}
\caption{Theoretical X-ray line ratios 3s/3d as a function of electron
temperature, or electron beam energy E = kT,
are compared with observations from the corona of solar-type star
Capella at
$\sim$~5--6\,MK from Chandra (open blue circle) \protect\cite{can00} and
XMM-Newton (filled blue circle)
\protect\cite{aud01}, from solar coronae at $T_{\rm m} \sim$~4\,MK (red
open circles) \protect\cite{hut76},
and from the LLNL and NIST EBIT experiments (filled and open green
circles respectively) \protect\cite{bei02,lam00}).
See text for more discussions about the details of the 3s/3d problem.
The scattered data for 3s/3d line ratios with
labels 1--14 are for various types of observed values from solar,
stellar and disk coronae \protect\cite{mck80,huen01,ness03,
mewe01,brink00,raa02,phi97,kahn01,xu02}.
\label{3s3d}
}
\end{minipage}
\end{figure*}

The main results of the calculations and modeling
of the 3s/3d line ratio are summarised in Fig.\,\ref{3s3d}.
The black arrow marks $T_{\rm M}\sim4 \times 10^6$ K, the
value of \Fe17\ maximum abundance in coronal equilibrium. The red arrows mark
different highest target thresholds that may be included in CC calculations.
Although this study is based on one of the largest and most computer
intensive BPRM close coupling
calculations, we could only include target levels up to $n=4$.
Comparison with the two experiments shows
that NIST EBIT data mimic our calculation assuming a Maxwellian averated
EDF (MA),
and LLNL data are in basic agreement with the GA results.
However, there is some discrepancy with the LLNL results
at the higher temperature end. The target levels with 
$n \geq 5$ are currently neglected in the
89CC BPRM calculations. This limitation could result in some
cascade contribution to
the 3s/3d X-ray line ratios \cite{lie99}, as well as some missing
resonance structures.
We have used an EBIT electron beam distribution with FWHM 
(full width at half maximum) of 30 eV in our GA modeling.

Fig.\,\ref{3s3d} shows two sets of 3s/3d intensity ratios 
as functions of electron temperature or electron beam energy.
The first set of calculations using MA collision strengths in 
the 89-level CR model appear as the black curve, calculated at
two electron densities 10$^{13}$ cm$^{-3}$ (black solid line) 
and 10$^9$ cm$^{-3}$ (black dotted line); dependence upon electron density 
is negligible.
The second set of data are shown by the blue solid curve with 
3s/3d ratios using GA collision strengths with a FWHM of 30 eV
(blue dotted curve is obtained with a 50 eV convolution which
gives essentially the same results); the further calculation with
a 5 eV convolution is also shown by the red solid curve.
The oscillations in the 3s/3d values in
the GA model are clearly seen. They are due to the distribution,
strength, and the
density of resonances in the collision strengths in different energy
regions as sampled by the convolution function centered around a 30 eV
or 5 eV FWHM Gaussian. This oscillation behavior is crucial in
explaining accurately the energy (temperature) dependence of 3s/3d ratios
in the EBIT experiments.
All previous calculations either fail to get the correct 3s/3d line ratios, or
fail to obtain their right temperature dependence.
The observed 3s/3d ratios from the solar corona are marked with 
red circles.
The open and filled blue circles with error bars mark in particular 
the coronal spectra of the solar-type star
Capella, observed from Chandra and XMM-Newton respectively \cite{can00,aud01}.
The LLNL EBIT experimental values are shown as green filled circles with
error bars, and the NIST EBIT experimental values as
two black open circles (without polarization corrections \cite{lam00})
or two green open circles (with polarization corrections \cite{gil05}).
N.\,B.\ NIST data in \cite{lam00} are reported without polarization
corrections while all the LLNL data in \cite{bei02} are reported with
polarization corrections. Also shown for comparison are the very scattered
3s/3d ratios from the long history of various observations of solar, stellar, 
and disk coronae (corresponding references are given in the caption of
Fig. 2).

The 3s/3d problem arises mainly from the factor two difference in the two 
EBIT experiments at the lower energy, and partly from the large scatter in
other observations. This problem is more serious than the 3C/3D problem, as 
mentioned above, because the two EBIT group are in good agreement for the
3C/3D ratio. Also, the 3s/3d ratio
is complicated by a complex line formation mechanism in the three 
3s X-ray lines. Both, the cascade effects from high-lying levels
and resonance effects, are important and need to be calculated accurately.
%in an extended (and computationally intensive) model.
First of all, we need to understand the difference in 
EBIT experimental values. We note that
the 3s/3d values from EBIT measurements need to be interpreted with care
in order to compare directly with astronomical observations because of
different plasma conditions and/or different EDF.
This point does not bear strongly on the 3C/3D problem, but for the 
3s/3d problem it is critical in order to derive a correct
interpretation. For example, a recent work  \cite{bei04} 
reports {\it similar} values for
the line ratio 3s/3C (which differs slightly from the ratio 3s/3d in the
denominator) from two different sources, the LLNL EBIT and the tokamak
Princeton Large Torus \cite{bei04}. However, this kind of agreement
as detailed in Fig.\ 7 of \cite{bei04} is most likely fortuitous.
This difference in magnitude for GA results, decreasing monotonically with
temperature over extended ranges but attenuated by oscillations due to
resonances in specific energy regions, is clearly seen in Fig.\,\ref{3s3d} and
is the general feature established in this work.
At lower energies the effect of resonances belonging to
low-lying levels is more pronounced, and decreases with successively higher
levels. This asymmetry in the distribution of resonance strengths with energy
is responsible for the overall behavior of the line ratio in
Fig.\,\ref{3s3d}.

Fig.\,\ref{3s3d} apparently shows that the NIST data at the lower energy
agrees better with the primary MA calculations, although the EDF inside both
the NIST and the LLNL EBITs is quite different from a Maxwellian. The
dichotomy inherent in these results reveals the importance of precise
determination of experimental beam energies and widths. The NIST data point
at the lower energy could possibly correspond to the minimum in the fundamental
oscillations of the 3s/3d line ratio, with a narrower EDF than the LLNL EBIT
\cite{gil05}. We further illustrate this by calculating a GA 3s/3d line ratio
with a FWHM of 5 eV as shown in the oscillating red solid curve in
Fig.\,\ref{3s3d}. We find that indeed the NIST data (the green open circle
with polarization correction) at the lower energy is located very close to a
minimum in this red curve. The NIST data point (with polarization correction)
at the higher energy is in agreement with our GA result. While at present the
two sets of experimental data do not fully trace the theoretically predicted
oscillations, the present results show that a much more detailed investigation
of the fundamental nature of the measurements is required in order to reconcile
the two sets of EBIT measurements, assuming that both are of claimed accuracy.
Accurate measurements of beam widths are crucial, but to our knowledge have not
been accurately determined or reported. The EDF, in turn, bears on physical
effects such as space charge potential, beam currents, polarization corrections
etc.\ \cite{gil05}. Nevertheless, the theoretical sets of 3s/3d ratios clearly
demonstrate that they are sensitive to the plasma conditions in the source
(c.\,f.\,\cite{bei02,bei04}). In order to interpret additional experimental
information on these issues, the theoretical results will also need to be
extended to other EDF to elicit a more complete description of resulting
variations in line intensities.

The large-scale relativistic CC calculations for EIE
of X-ray lines in \Fe17 \cite{che02,che03}, and
the study of different electron distribution functions in plasmas {\it via\/}
CR models, appear to have resolved long-standing 
problems. % as manifest in the 3C/3D and 3s/3d line ratios. 
The relativistic atomic calculations of \Fe17 and the
line ratios among its diagnostic X-ray lines
are in agreement with two independent measurements on EBIT
\cite{bei02,lam00}.
It follows that the
line ratios from the present 89CC CR model may be further applied to interpret
X-ray spectra from a variety of X-ray sources observed from current
space observatories, Chandra and XMM-Newton, and the upcoming 
next generation ones with greater spectroscopic capabilities,
such as Constellation-X. 
Towards this goal additional
calculations are in progress to
 calculate level-specific and total
recombination rates from from Fe\ {\sc XVIII} to \Fe17, including
radiative and dielectronic recombination
in a unified scheme \cite{np04}, and photoionization
rates from Fe\ {\small XVI} to Fe\ {\small XVII}.
However, astrophysical environments and plasma
conditions may be different from those in EBIT plasmas.
It is our goal to apply our `calibrated' atomic data to interpret
observed stellar and cosmic X-ray spectra
of coronal or collisionally ionized
plasmas such as stellar coronae, and photoionized or hybrid plasmas in, 
for example, from 
active galactic nuclei (AGN), quasars, disk coronae etc.

In summary, the new results for \Fe17 structure, collision dynamics,
and spectral diagnostics
reveal not only the effect of resonances and cascades, but also present
a challenge for the study of
source-specific effects of electron distributions and other conditions 
prevalent in EBIT, tokamak, laser-produced, and astrophysical plasmas.
The present GA and MA averaged CR models should
offer possible solutions to obtain more precise
information on the physical parameters in the beam and the trap in EBIT
experiments.
We demonstrate that 3s/3d X-ray line ratios are not a universal constant
in an optically thin plasma excited by electron impact, but instead depends
sensitively on both the characteristic electron energy (temperature) and
the detailed distribution of the electron energies about that characteristic
value.
Furthermore, the present results should be of general
astrophysical importance for objects in a variety of equilibrium or
non-equilibrium conditions.
A direct physical consequence of this work is
that the departure from a Maxwellian should manifest itself in, and be used
as a diagnostics of, particle distributions in plasmas.
In photoionization equilibrium 
(as opposed to coronal) we also need level-specific 
recombination cross sections and rates in order to compute emission
line intensities (work in progress). Finally, the results
in this Letter should be applicable to 
other neon-like ions of importance in X-ray
lasers \cite{ros85}, with
population inversion among the upper $n$=3 levels due to EIE and/or
electron-ion recombination.

\section{Conclusion}

 The main conclusions of the present study are:

 $\bullet$ Line intensities and ratios of the 3s and 3d line-complexes at
 $\sim 17 \AA$ and $\sim 15 \AA$ depend on electron 
 distribution functions. The 3s/3d ratio
 can be a potential diagnostics of temperature, particularly for
 T $\leq T_m$ in
 astrophysical sources with Maxellian EDF's, and of
 energy, particularly at $E > kT_m$ in non-Maxwellian plasmas such as
 laboratory sources.

 $\bullet$ The Maxwellian and Gaussian simulations of \Fe17 X-ray line ratios  
 provide an indication of diffrerences in the conditions in the
 interaction regions of Electron-Beam-Ion-Traps; however, more
 experiments are needed to fully understand the physical factors
 involved.

 $\bullet$ X-ray line ratios are a potential diagnostic of
 non-maxwellian astrophysical and laboratory sourcs, 
 such as accretion discs, X-ray flares and bursts, 
 and magnetically or inertially confined fusion plasmas. 

\section*{Acknowledgments}

We would like to thank Drs.\ John Gillaspy and Martin Laming
for discussions concerning the details of EBIT measurements.
This work
was partially supported by the NASA Astrophysical Theory Program.
The computational work was carried out on the Cray SV1 at the Ohio Supercomputer Center in Columbus Ohio.

\label{lastpage}

\end{document}